\documentclass[letters, fleqn, usenatbib]{mnras}
\usepackage{amssymb,amsmath}
\usepackage{graphicx}
\usepackage{xcolor}
\usepackage{multirow}
\usepackage[T1]{fontenc}
\usepackage{ae,aecompl}
\usepackage{newtxtext,newtxmath}

\def\msig{$M_{\rm BH}- \sigma$\ }

\title[Central TDE in NGC 1097]
{Modeling the flare in NGC 1097 from 1991 to 2004 as a tidal disruption event}
\author[Zhang]
{Xue-Guang Zhang$^1$\thanks{Corresponding author Email: \href{mailto:aexueguang@qq.com}
{aexueguang@qq.com}}\\
$^{\color{blue}1*}$ School of Physical Science and Technology, GuangXi University, 
	No. 100, Daxue Road, 530004, Nanning, P. R. China}

\pubyear{2022}

\begin{document}

\label{firstpage}

\pagerange{\pageref{firstpage}--\pageref{lastpage}}

\maketitle

\begin{abstract} 
	In the Letter, interesting evidence is reported to support a central tidal 
disruption event (TDE) in the known AGN NGC 1097. Considering the motivations of TDE 
as one probable origination of emission materials of double-peaked broad emission 
lines and also as one probable explanation to changing-look AGN, it is interesting 
to check whether are there clues to support a TDE in NGC 1097, not only a changing-look 
AGN but also an AGN with double-peaked broad emission lines. Under the assumption 
that the onset of broad H$\alpha$ emission was due to a TDE, the 13years-long 
(1991-2004) variability of double-peaked broad H$\alpha$ line flux in NGC 1097 can be 
well predicted by theoretical TDE model, with a $(1-1.5){\rm M_\odot}$ main-sequence 
star tidally disrupted by the central BH with TDE model determined mass about 
$(5-8)\times10^7{\rm M_\odot}$. The results provide interesting evidence to not 
only support TDE-related origin of double-peaked broad line emission materials but 
also support TDE as an accepted physical explanation to physical properties of 
changing-look AGN. 
\end{abstract}


\section{Introduction}

	TDEs (Tidal Disruption Events) have been well studied in detail for more than four 
decades \citep{re88, lu97, ce12, gs12, gr13, gm14, ko15, wy18, tc19}, with accreting fallback 
debris from stars tidally disrupted by central black holes (BHs) leading to apparent 
time-dependent variability. Based on TDE expected variability properties, there are more than 
100 TDE candidates detected and reported in the literature (see the collected TDE candidates 
listed in \url{https://tde.space/}), strongly supporting the idea that TDEs can be used to 
locate massive BHs and accreting BH systems. More recent reviews on theoretical simulations 
on TDEs can be found in \citet{st18} and on detected TDE candidates can be found in \citet{gs21}. 
More recent large samples of dozens of TDE candidates can be found in the \citet{vg21} from 
the First Half of ZTF (Zwicky Transient Facility) Survey observations and in \citet{sg21} 
from the SRG all-sky survey observations.

	Along with studying on TDEs, some special spectroscopic features have been considered 
to be tightly related to TDEs, such as double-peaked broad emission lines, see the discussions 
in \citet{el95} for AGN with double-peaked broad emission lines (hereafter, double-peaked AGN). 
More detailed discussions on variability properties of double-peaked broad emission lines in 
TDE candidates can be found in SDSS J0159 in \citet{md15, zh21}, in ASASSN-14li in \citet{ht16}, 
in PTF09djl in \citet{lz17}, in PS18kh in \citet{ht19}, in AT2018hyz in \citet{sn20, hf20}, 
etc.. Therefore, it is interesting to check whether are there double-peaked AGN as host galaxies 
of TDE candidates. Moreover, TDEs could be probably related to or probably detected in 
changing-look AGN (AGN with type transitioned between Type 1 and Type 2), such as the detailed 
results in the changing-look AGN SDSS J0159 in \citet{md15, ld15} and more detailed discussions 
on a sample of changing-look AGN in \citet{yw18} and in the bluest changing-look quasar in 
\citet{zh21b}. Therefore, TDE candidates could be more preferred in changing-look double-peaked AGN.

	Fortunately, there is an AGN, the low luminosity AGN NGC 1097, which has been 
classified as a double-peaked AGN and also as a changing-look AGN. Changing-look properties 
of NGC 1097 can be confirmed by the following spectroscopic features: no broad emission lines 
before 1990 as reported in \citet{wn86} but apparently double-peaked broad emission lines in 
the 1990s and in early 2000s as discussed in \citet{sb93, sn03}. Moreover, \citet{kk12} have 
studied properties of central regions of NGC 1097 through near-infrared spectrum and reported 
that there are no any evidence for nuclear activity in NGC 1097\ in Jul. 18th, 2008, which can 
be well applied to discuss physical models of flare in NGC 1097. And, the apparently detected 
double-peaked broad H$\alpha$ lead NGC 1097 to be clearly classified as a double-peaked AGN, 
although NGC 1097 has central continuum luminosity at 5100\AA~ around $10^{40}{\rm erg/s}$ 
(see Fig.~3\ in \citet{sn05}) much lower than common values around $10^{44}{\rm erg/s}$ of 
broad line AGN \citep{sh11}.

	In the Letter, variability properties of NGC 1097 are well studied to check whether 
is there strong evidence to support a central TDE in NGC 1097. Section 2 presents our main 
results on variability of broad H$\alpha$ line flux from 1991 to 2004, and necessary 
discussions. Section 3 gives our final conclusions. And in the Letter, we have adopted the 
cosmological parameters of $H_{0}=70{\rm km\cdot s}^{-1}{\rm Mpc}^{-1}$, $\Omega_{\Lambda}=0.7$ 
and $\Omega_{\rm m}=0.3$.

\section{Long-term Variability of double-peaked broad H$\alpha$ line flux from 1991 to 2004} 

\begin{figure}                                                                    
\centering\includegraphics[width = 8cm,height=5cm]{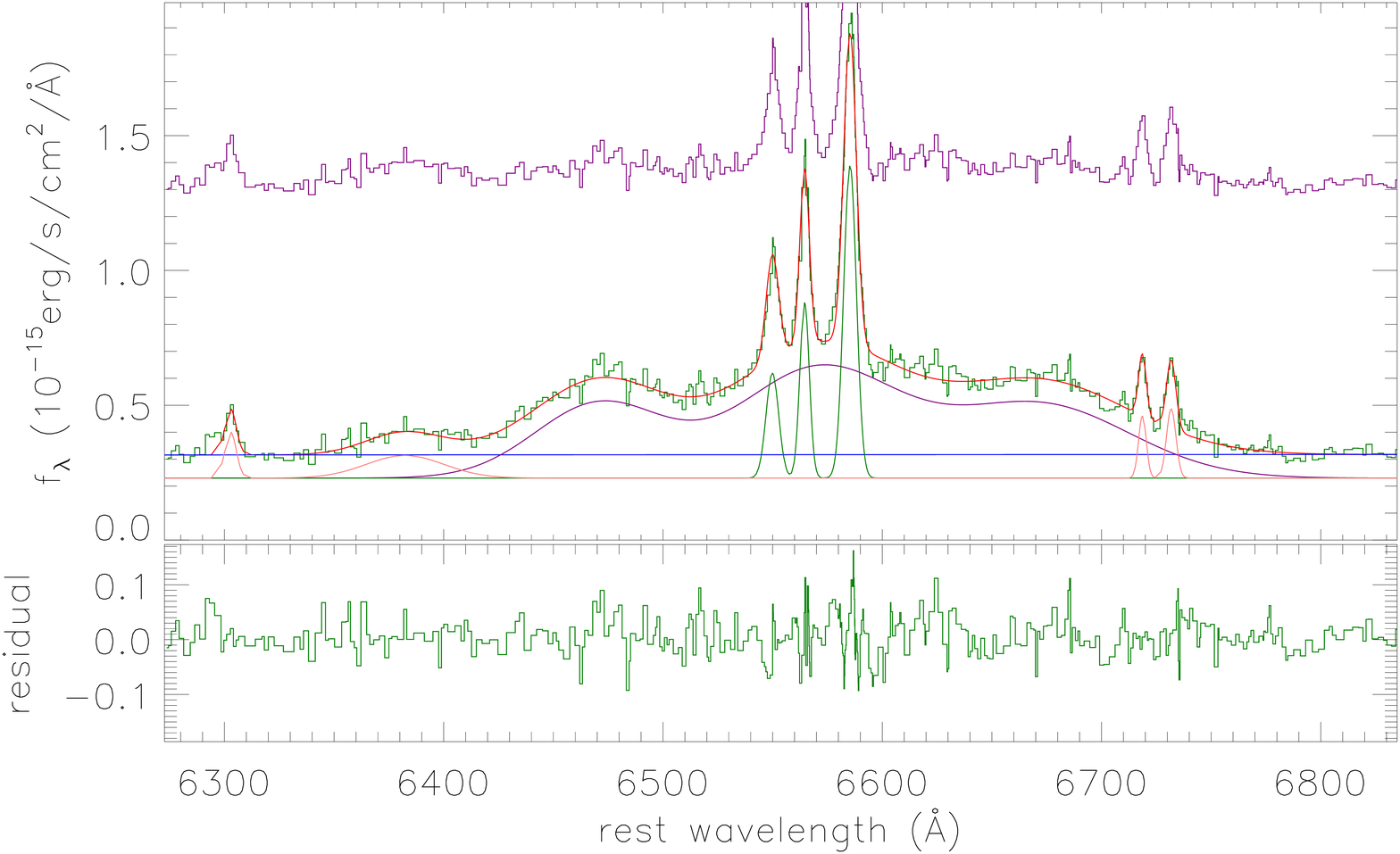}                
\caption{The best fitting results (top panel) and the corresponding residuals (bottom panel) 
(spectrum minus the best fitting results) to the emission lines around H$\alpha$ of NGC 1097 
observed in 2004 by multiple Gaussian functions. In top panel, solid lines in dark green and 
in red represent the observed spectrum and the corresponding best fitting results, solid lines 
in purple, blue, pink and dark green show the determined broad H$\alpha$, the determined power 
law component, the determined [O~{\sc i}] and [S~{\sc ii}] doublets, the determined narrow 
H$\alpha$ and [N~{\sc ii}] doublet, respectively. In top panel, solid line in purple represents 
the expected spectrum plus 1\ in 2008 through the spectrum in 2004 with the broad component 
weakened by TDE model predicted factor of 3.5.}
\label{line}                                                                     
\end{figure}

	Besides spectroscopic results in \citet{sn03} from 1991 to 2001, another spectroscopic 
results can be collected from the HST (Hubble Space Telescope) mission (ID:8684, PI: Dr. 
Eracleous) in 2004 with broad H$\alpha$ line flux about 
$F_{b,obs}\sim(85.6\pm7.1)\times{\rm 10^{-15}erg/s/cm^2}$. Here, three broad Gaussian 
components are applied to describe the broad H$\alpha$ and the other seven narrow 
Gaussian components are applied to describe the narrow H$\alpha$, [O~{\sc i}]$\lambda6300, 
6363$\AA, [N~{\sc ii}]$\lambda6548, 6583$\AA~ and [S~{\sc ii}]$\lambda6716, 6731$\AA\ doublets. 
Through the Levenberg-Marquardt least-squares minimization technique, the emission components 
can be well determined and shown in Fig.~\ref{line}.

	As discussed in \citet{sn03}, in order to correct effects of aperture sizes on optical 
spectra with different instruments, total line intensity of narrow H$\alpha$, [N~{\sc ii}] and 
[S~{\sc ii}] doublets are applied to determined a scaling factor. Here, the determined total 
narrow line flux in 2004 is $172\times{\rm 10^{-15}erg/s/cm^2}$. Meanwhile, the total 
narrow line flux in Nov. 1991 is $93\times{\rm 10^{-15}erg/s/cm^2}$ as shown in \citet{sb93}. 
Therefore, the scaling factor for the spectrum in 2004 is about $(93\times1.12)/172\sim0.61$ 
with $1.12$ as the determined scaling factor for the spectrum in Nov. 1991 relative to the 
spectrum in Jan. 1996 (the standard spectrum in \citet{sn03}). After corrections of aperture 
effects, the finally accepted broad H$\alpha$ line flux in 2004 is 
$\sim0.61\times F_{b, obs}=(51.8\pm4.3)\times{\rm 10^{-15}erg/s/cm^2}$. Moreover, as shown in 
Fig.~\ref{line}, there is a broad component around [O~{\sc i}]$\lambda6363$\AA. It is not clear 
the broad component (line flux about $(3.7\pm0.5)\times{\rm 10^{-15}erg/s/cm^2}$) is from the 
[O~{\sc i}] or from the broad H$\alpha$, however, the lower line flux of the broad component has 
few effects on the following results. Therefore, there are no further discussions on the 
weak broad component around [O~{\sc i}]$\lambda6363$\AA.

\begin{figure}
\centering\includegraphics[width = 8cm,height=5cm]{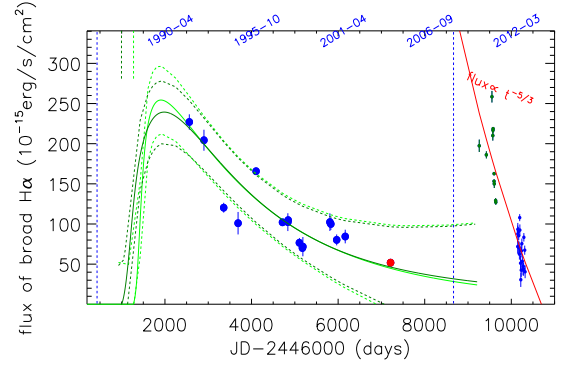}
\caption{Best descriptions to the variability of broad H$\alpha$ line flux by the TDE model. 
Large solid blue circles plus error bars are the data points reported in \citet{sn03}, large 
solid red circle plus error bar represents the calculated value in 2004\ in the Letter. Solid 
and dashed lines in green and in dark green represent the best descriptions determined by the 
TDE model with $\gamma=4/3$ and with $\gamma=5/3$ to the variability from 1991 to 2004, and 
the corresponding 90\% confidence bands through the F-test technique, respectively. Small 
circles plus error bars in dark green and in blue represent the data points collected from 
\citet{ss12} and from \citet{ss15}, respectively. Solid red line shows the $t^{-5/3}$ variability 
trend which can be applied to roughly describe the variability from 2010 to 2013. In the panel, 
the two vertical blue lines from left to right mark the positions relative to Jan. 1th, 1986 
(corresponding to no broad H$\alpha$) and Jul. 18th, 2008 (no apparent central activity through 
properties of near-infrared spectrum). In top region, the corresponding date information of 
year-month are marked for the JD-2446000=2000, 4000, 6000, 8000, 10000, respectively, and vertical 
dashed lines in green and in dark green mark the starting time for the TDE with $\gamma=4/3$ and 
$\gamma=5/3$, respectively.
}
\label{tde}
\end{figure}

	The broad H$\alpha$ line flux in \citet{sn03} and the new calculated broad H$\alpha$ 
line flux in 2004 can lead to the long-term variability over ten years from 1991 to 2004 shown 
in Fig.~\ref{tde}. It is interesting that the flux variability of broad H$\alpha$ has a 
systematically time-dependent declined trend, similar as TDE expected properties. Then, the 
more recent theoretical TDE model well discussed in \citet{gr13, gm14, mg19} can be applied 
as follows to check whether a TDE model can be applied to describe the long-term variability 
of broad H$\alpha$ flux.  
   
	There are public codes on the theoretical TDE model, such as the TDEFIT and the 
MOSFIT. Here, based on the more recent discussions in \citet{mg19}, the theoretical TDE model 
can be applied by the following four steps to model variability of NGC 1097, similar as what 
we have done in \citet{zh22} to describe the X-ray variability in the TDE candidate {\it Swift} 
J2058.4+0516 with relativistic jet.

	First, standard templates of viscous-delayed accretion rates $\dot{M}_{at}$ are created by, 
\begin{equation}
\begin{split}
\dot{M}_{at}~&=~\frac{exp(-t/T_{v})}{T_{v}}\int_{0}^{t}exp(t'/T_{v})\dot{M}_{fbt}dt' \\
\dot{M}_{fbt}~&=~dm/de~\times~de/dt~=~dm/de~\times~\frac{(2~\pi~G~M_{\rm BH})^{2/3}}{3~t^{5/3}}
\end{split}
\end{equation}
with $dm/de$ as the TDEFIT provided distributions of debris mass $dm$ as a function of specific 
binding energy $e$ after a star is disrupted, $M_{\rm BH}$ as central BH mass, and $\dot{M}_{fbt}$ 
as TDEFIT and MOSFIT provided templates of fallback material rates for standard cases with central 
BH of $M_{\rm BH}=10^6{\rm M_\odot}$ and disrupted main-sequence star of $M_{*}=1{\rm M_\odot}$ 
and with a grid of the listed impact parameters $\beta_{temp}$ in \citet{gr13}, and $T_{v}$ as 
the viscous time after considering the viscous delay effects as discussed in \citet{gr13, mg19}. 
Here, a grid of 31 $\log(T_{v, temp}/{\rm years})$ range from -3 to 0 are applied to create 
templates of $\dot{M}_{at}$ for each impact parameter. Finally, templates of $\dot{M}_{at}$ include 
736 (640) time-dependent viscous-delayed accretion rates for 31 different $T_{v}$ of each 23 
(20) impact parameters for the main-sequence star with polytropic index $\gamma$ of 4/3 (5/3).

	Second, simple linear interpolations are applied to determine accretion rates 
$\dot{M}_{a}(T_{v}, \beta)$ for TDEs with input model parameters of $\beta$ and $T_{v}$ 
different from the list values in $\beta_{temp}$ and in $T_{v, temp}$. Assuming that 
$\beta_1$, $\beta_2$ in the $\beta_{temp}$ are the two values nearer to the input $\beta$ and 
$T_{v1}$, $T_{v2}$ in the $T_{v,temp}$ are the two values nearer to the input $T_{v}$, 
the expected $\dot{M}_{a}(T_{v}, \beta)$ can be estimated by
\begin{equation}
\begin{split}
\dot{M}_{a}(T_{v}, \beta_{1}) &= \dot{M}_{at}(T_{v1}, \beta_1) + \\ 
	&\frac{T_{v}-T_{v1}}{T_{v2}-T_{v1}}(\dot{M}_{at}(T_{v2}, \beta_1)
	- \dot{M}_{at}(T_{v1}, \beta_1))\\
\dot{M}_{a}(T_{v}, \beta_2) &= \dot{M}_{at}(T_{v1}, \beta_2) + \\
	&\frac{T_{v}-T_{v1}}{T_{v2}-T_{v1}}(\dot{M}_{at}(T_{v2}, \beta_2)
	- \dot{M}_{at}(T_{v1}, \beta_2)) \\
\dot{M}_{a}(T_{v}, \beta) &= \dot{M}_{a}(T_{v}, \beta_1) + 
	\frac{\beta-\beta_1}{\beta_2-\beta_1}(\dot{M}_{a}(T_{v}, \beta_2) - 
	\dot{M}_{a}(T_{v}, \beta_1))
\end{split}
\end{equation}

        Third, for TDEs with input $M_{\rm BH}$ and $M_{*}$ different from $10^6{\rm M_\odot}$ 
and $1{\rm M_\odot}$, as discussed in \citet{gr13, mg19}, actual viscous-delayed accretion rates 
$\dot{M}$ and the corresponding time information are created by the following scaling rations 
applied with input BH mass, mass and radius of the disrupted main-sequence star,
\begin{equation}
\begin{split}
&\dot{M} = M_{\rm BH,6}^{-0.5}\times M_{\star}^2\times
	R_{\star}^{-1.5}\times\dot{M}_{a}(T_{v}, \beta) \\
&t = (1+z)\times M_{\rm BH,6}^{0.5}\times M_{\star}^{-1}\times
	R_{\star}^{1.5} \times t_{a}(T_{v}, \beta)
\end{split}
\end{equation},
where $M_{\rm BH,6}$, $M_{\star}$, $R_{\star}$ and $z$ represent central BH mass in unit of 
${\rm 10^6M_\odot}$, stellar mass in unit of ${\rm M_\odot}$, stellar radius in unit of 
${\rm R_{\odot}}$ and redshift of host galaxy of a TDE, respectively. And the mass-radius 
relation discussed in \citet{tp96} has been accepted in the Letter for main-sequence stars.

        Fourth, the time-dependent output emission spectrum in rest frame based on the TDE model 
expected accretion rate $\dot{M(t)}$ can be determined by the simple black-body photosphere model 
as discussed in \citet{gm14, mg19},
\begin{equation}
\begin{split}
&F_\lambda(t)=\frac{2\pi Gc^2}{\lambda^5}\frac{1}{exp(hc/(k\lambda T_p(t)))-1}(\frac{R_p(t)}{D})^2 \\
&R_p(t) = R_0\times a_p(\frac{\eta\dot{M(t)}c^2}{1.3\times10^{38}M_{\rm BH}/{\rm M_\odot}})^{l_p} \\
&T_p(t)=(\frac{\eta\dot{M(t)}c^2}{4\pi\sigma_{SB}R_p^2})^{1/4} \ \ \ \ \ 
	a_p = (G M_{\rm BH}\times (\frac{t_p}{\pi})^2)^{1/3}
\end{split}
\end{equation}
where $D$ means the distance to the earth calculated by redshift $z=0.00424$ (the redshift of NGC 
1097), $k$ is the Boltzmann constant, $T_p(t)$ and $R_p(t)$ represent the time-dependent effective 
temperature and radius of the photosphere, respectively, and $\eta$ is the energy transfer 
efficiency smaller than 0.4, $\sigma_{SB}$ is the Stefan-Boltzmann constant, and $t_p$ is the 
time information of the peak accretion. Then, based on the $F_{5100\textsc{\AA}}(t)$ in rest 
frame, time-dependent broad H$\alpha$ line fluxes $f_{H\alpha}(t)$ can be determined by
\begin{equation}
f_{H\alpha}(t)~=~F_{5100\textsc{\AA}}(t)~\times~K_{cl}
\end{equation}
with $K_{cl}$ as the intensity ratio of broad H$\alpha$ line flux to continuum emission intensity 
at 5100\AA~ in rest frame, through the strong correlation between continuum luminosity and broad 
H$\alpha$ line luminosity \citep{gh05, zh08}.

\begin{figure*}
\centering\includegraphics[width = 18cm,height=6cm]{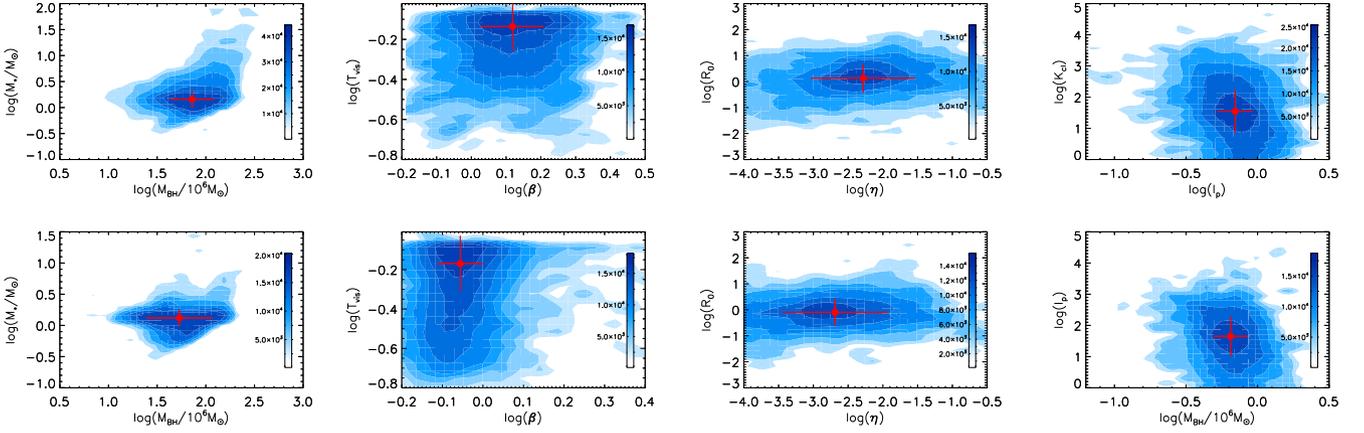}
\caption{Two-dimensional posterior distributions of the model parameters with $\gamma=4/3$ (top 
panels) and with $\gamma=5/3$ (bottom panels). In each panel, solid red circle plus error bars 
show the final accepted values and corresponding uncertainties of the model parameters. 
Contour levels with different colors represent different number densities, as 
shown in the color bar in each panel.}
\label{par}
\end{figure*}

	When the procedure above is applied to describe the observed variability of broad 
H$\alpha$ line flux, there is only one limitation that the determined tidal disruption radius 
$R_{\rm TDE}\propto(M_\star)^{-1/3}(M_{\rm BH})^{-2/3}R_\star$ to be be larger than event 
horizon of central BH $R_{\rm G}=2GM_{BH}/c^2$.

	Finally, the theoretical TDE model expected time dependent light curves $f_{H\alpha}(t)$ 
can be described by eight model parameters, central BH mass $\log(M_{\rm BH})$, stellar mass 
$\log(M_\star)$ (corresponding stellar radius $R_\star$ calculated by the mass-radius relation), 
energy transfer efficiency $\log(\eta)$, impact parameter $\log(\beta)$, viscous timescale 
$\log(T_{v})$, parameters $l_p$ and $R_0$ related to the black-body photosphere model, and ratio 
$K_{cl}$. Then, through the well-known maximum likelihood method combining with the Markov Chain 
Monte Carlo (MCMC) technique \citep{fh13}, the variability of broad H$\alpha$ line fluxes in NGC 
1097 can be well predicted and shown in Fig.~\ref{tde} from 1991 to 2004, accepted prior uniform 
distributions of the model parameters and the corresponding starting values listed in Table~1. 
Two-dimensional posterior distributions of the model parameters are shown in Fig.~\ref{par}. 
And the final accepted model parameters and corresponding uncertainties are also listed in Table~1. 
Here, the uncertainty of each parameter is determined by the half width at half maximum of 
distribution of each parameter. Based on the determined model parameters, the expected TDE is 
starting from JD=$2447275\pm140$ (around Apr. 24th, 1988) and from JD=$2446997\pm180$ (around 
Jul. 20th, 1988) for the model with $\gamma=4/3$ and $\gamma=5/3$, respectively. If considering 
the mass limitation of central disrupted stellar in \citet{gr13, mg19}: $M_\star<0.3{\rm M_\odot}$ 
or $M_\star>22{\rm M_\odot}$ for $\gamma=5/3$, the TDE model with $\gamma=4/3$ is preferred in 
NGC 1097. The determined model parameters have reasonable values, compared with the values for 
the other optical TDEs candidates in the literature, providing direct and interesting clues to 
support a central TDE in NGC 1097\ from 1991 to 20004.

\begin{table}
\caption{Parameters of TDE models for NGC 1097}
\begin{tabular}{lcrll}
\hline\hline
parameter & prior & p0 & value$^a$ & value$^b$ \\
\hline
$\log(M_{\rm BH,~6})$ & [-3, 3] & 1. & $1.86\pm0.22$ & $1.73\pm0.34$ \\
$\log(M_\star/M_\odot)$ & [-2, 1.7] & 0. & $0.16\pm0.17$ & $0.12\pm0.13$ \\
$\log(\beta)(4/3)$ & [-0.22, 0.6] & 0. & $0.12\pm0.09$ & \dots \\
$\log(\beta)(5/3)$ & [-0.3, 0.4] & 0. & \dots & $-0.06\pm0.06$ \\
$\log(T_{vis})$ & [-3, 0] & -1. & $-0.14\pm0.12$ & $-0.17\pm0.14$\\
$\log(\eta)$  & [-3, -0.4] & -1. & $-2.28\pm0.75$ & $-2.69\pm0.76$ \\
$\log(R_{0})$ &  [-3, 3] & 1. & $0.13\pm0.55$ & $-0.10\pm0.55$ \\
$\log(l_{p})$  & [-3, 0.6] & 0. & $-0.16\pm0.12$ & $-0.19\pm0.12$ \\
$\log(K_{cl})$  & [-3, 6] & 1. & $1.55\pm0.59$ & $1.65\pm0.64$ \\
\hline\hline
\end{tabular}
Notes: The first column shows the applied model parameters. The second column shows 
limitations of the prior uniform distribution of each model parameter. The third 
column with title "p0" lists starting value of each parameter. The fourth column 
with column title marked with $^a$ means the values of model parameters for the 
TDE model with $\gamma=4/3$. The fifth column with column title marked with $^b$ means 
the values of model parameters for the TDE model with $\gamma=5/3$.
\end{table}

	Then, it is interesting to check whether the TDE model predicted H$\alpha$ 
line flux is lower enough to be consistent with the reported none apparent AGN 
activity in 2008\ in \citet{kk12}. As shown in Fig.~\ref{tde}, the expected broad 
H$\alpha$ line flux in 2008 is about $25\times10^{-15}{\rm erg/s/cm^2}$, about 3.5 
times weaker than that observed in 2004. Now, it is interesting to check whether so weak 
broad H$\alpha$ component can be well detected in spectrum. As an oversimplified example, 
based on the observed spectrum in 2004 with the broad component weakened by a factor 
of 3.5 (similar values for the TDE model with different $\gamma$), the expected 
spectrum in 2008 is shown in Fig.~\ref{tde}: no apparent broad H$\alpha$, to be 
consistent with the reported none-activity in NGC 1097\ in 2008\ in \citet{kk12}. In 
other words, the none apparent activity in 2008 reported in \citet{kk12} could be 
simply expected, under the assumption of a central TDE in NGC 1097. Therefore, the 
determined TDE model is reasonable.

	Based on the best descriptions shown in Fig.~\ref{tde}, the variability from 1991 to 2004\ 
in NGC 1097 can be well predicted by theoretical TDE model, under the assumption of double-peaked 
broad H$\alpha$ emissions related to a TDE. However, there is an interesting question whether is 
there further evidence to support the emission materials of the double-peaked broad emission lines 
related to the central TDE. As discussed in \citet{sn03}, the double-peaked features provide the 
corresponding emission regions with inner radius about $225R_{\rm G}$ and outer radius about 
$800R_{\rm G}$ (here, $R_{\rm G}$ is two times of the unit used in \citet{sn03}). If the 
double-peaked broad H$\alpha$ were related to the central TDE, we could expect the regions of 
accreting debris in the TDE could cover the emission regions for the double-peaked broad H$\alpha$. 
Based on the model parameters, the distance of accreting debris as discussed in \citep{gm14} can 
be roughly estimated by $R_o\sim2\times(\frac{GM_{\rm BH}\times t^2}{\pi^2})^{1/3}$, leading to 
$R_o\sim3200{\rm R_G}$ in 2000 (where $t$ is about 3400days), indicating the accreting debris 
totally cover the expected emission regions of the double-peaked broad H$\alpha$ in NGC 1097. 
Therefore, it is plausible to accept that the double-peaked broad H$\alpha$ are from accreting 
debris in the central TDE in NGC 1097. Furthermore, through the well measured stellar velocity 
dispersion about $190{\rm km/s}$ in NGC 1097\ in \citet{le06} and discussed in \citet{sn03, oi15}, 
the central BH mass of NGC 1097 has been reported to be around $10^8{\rm M_\odot}$ through the 
well-known \msig~ \citep{fm00, ge00, kh13, bb17, bt21}, simply consistent with the TDE model 
determined BH mass after considering the uncertainties, providing further evidence to support 
the central TDE in NGC 1097. 

	Before end of the section, three further points are noted. First, besides the long-term 
variability of double-peaked broad H$\alpha$ over ten years from 1991 to 2004, there are apparent 
variability of double-peaked broad H$\alpha$ from 2010 to 2013, as discussed in \citet{ss12, ss15}, 
which are shown as small circles in Fig.~\ref{tde}. Considering the long-term variability from 
1991 to 2013\ in NGC 1097 shown in Fig.~\ref{tde}, the variability pattern is quite like the case 
discussed in \citet{cm15, gk15} in the known changing-look AGN IC 3599. Different models have 
been proposed to explain the multiple flares. \citet{cm15} have shown that a tidal stripping in 
partial TDE can lead to repeated tidal disruption flares. However, \citet{gk15} have shown that 
the second flare in IC 3599 should be related to accretion disk variability. \citet{ml15} 
have shown that binary stars disrupted by central massive BH (double TDE) can lead to repeated 
flares. There are so-far no confirmed physical origin of repeated flares in IC 3599, neither in NGC 
1097. However, based on special properties of NGC 1097, further discussions can be given on the 
proposed models. \citet{ml15} have shown that the majority of double-star disruptions 
produce two flares close enough in time that they overlap. As NGC 1097 was apparently inactive in 
2008 \citep{kk12}, between the two peaks of H$\alpha$ flux, then this scenario in \citet{ml15} 
should be disfavoured. Probably, the second flare in NGC 1097 should be related to an independent 
TDE, because the TDE expected $t^{-5/3}$ (shown as solid red line in Fig.~\ref{tde}) can be roughly 
applied to describe the variability from 2011 to 2013.

	Second, as the discussed variability around 2010\ in \citet{ss12} (time duration less than 
400days), there is an interesting plateau phase (shown in Fig.~5\ in \citet{ss12}), which can not 
expected by standard TDE model. The interesting variability properties could including effects of 
accretion disk variability. Therefore, long-term variabilities with scale about ten 
years could be simply expected by standard TDE model, however, short-term variability with time 
scale about hundreds of days should be related to central accretion disk variability. 
Therefore, the long-term variability from 1991 to 2004 is mainly considered in the Letter, due to 
well predicted by the standard TDE model.

	Third, as the shown best fitting results in Fig.~\ref{tde} from 1991 to 2004, a decay in 
flux of a factor 5\ in 10 years is slow for standard TDEs. However, after considering viscous delay 
effects, the decaying trend should be quite flatter than $t^{-5/3}$. Similar flat decay can be 
found in some other reported TDEs, such as the PS1-11af, TDE2, etc. shown in Fig.~1\ 
in \citet{mg19} with magnitude decay about 1.5-2mags (flux decay factor about 4-6) from 
the time with peak intensity to late times. Therefore, the slower decay trend in NGC 1097 
can be well accepted. Certainly, as reported and discussed in \citet{gs08, gs09} and simple 
descriptions in \citet{gs21}, quite flatter decaying trend could be also found in TDE 
expected light curves at late stages, such as in the TDE candidate D3-13 with index 
$t^{-0.82}$ at late stages. Therefore, it is interesting to consider whether the flatter 
decaying trend in NGC 1097 is due to catching the event in the late stages. If accepted the 
variability from 1991 to 2004 from TDE expected variability at late stages, expected starting 
time of the TDE should be earlier than 1986, under simply considering the flatter decaying 
trend at late stages having similar (or even two times) time duration as the steeper 
decaying trend at early stages around the peak. However, there are no broad H$\alpha$ 
emissions in 1986, indicating the consideration of catching the event in the late stages 
should be not favoured to explain the flatter decaying trend in NGC 1097. Moreover, based 
on the TDE model determined parameters, the peak accretion rate is estimated to be 
$\dot{M}\sim0.02{\rm M_\odot/year}$, and the total accreted mass during the 13years-long 
flare is estimated to be $0.2{\rm M_\odot}$. Considering the determined energy efficiency 
about $\log(\eta)\sim-2.28$, the corresponding observed peak bolometric luminosity is about 
$\eta\times\dot{M}c^2\sim6\times10^{42}{\rm erg/s}$. However, after considering the probable 
intrinsic reddening effects with $A_{V}=3mag$ as discussed in \citet{sn05}, the intrinsic 
peak bolometric luminosity is about $9\times10^{43}{\rm erg/s}$ well consistent with the 
peak bolometric luminosities around $10^{44}{\rm erg/s}$ of the other TDE candidates as 
discussed in \citet{mg19}, to support the central TDE in NGC 1097.

\section{Conclusions}

	Finally, we give our main conclusions as follows. Under the assumption of double-peaked 
broad emission lines coming from TDE related debris, theoretical TDE model with a 
$1-1.5{\rm M_\odot}$ main-sequence star disrupted by the central BH with TDE model determined 
mass about $(5-8)\times10^7{\rm M_\odot}$ can be well applied to predict the 13years-long 
(1991-2004) variability of double-peaked broad H$\alpha$ line fluxes in the well-known low 
luminosity AGN NGC 1097, providing interesting evidence to support a central TDE in NGC 1097 
which has been classified as both a changing-look AGN and a double-peaked AGN.  Therefore, 
NGC 1097 is so-far the best candidate to support the TDE related origin of double-peaked broad 
Balmer emission materials and to support TDE as the physical explanation to properties of 
changing-look AGN.

\section*{Acknowledgements}
Zhang gratefully acknowledges the anonymous referee for giving us constructive 
comments and suggestions greatly improving our paper. Zhang gratefully acknowledges the 
great comments and suggestions from Prof. T. Storchi-Bergmann, and acknowledges the funding 
support from NSFC-12173020. This letter has made use of the NASA/IPAC Extragalactic Database 
(NED) operated by the Jet Propulsion Laboratory, Caltech, and the observation data with the 
NASA/ESA Hubble Space Telescope obtained from the Space Telescope Science Institute, and the 
public codes of TDEFIT (\url{https://github.com/guillochon/tdefit}) and MOSFIT 
(\url{https://github.com/guillochon/mosfit}).

\section*{Data Availability}
The data underlying this article will be shared on request to the corresponding author
(\href{mailto:aexueguang@qq.com}{aexueguang@qq.com}).

\label{lastpage}
\end{document}